\documentclass[11pt ]{article}
\RequirePackage[T1]{fontenc}
\usepackage[english]{babel}
\RequirePackage{amsthm,amsmath}
\usepackage[a4paper]{geometry}
\usepackage{setspace}
\usepackage[usenames, dvipsnames]{color}
\usepackage{graphicx}
\usepackage[outdir=./art]{epstopdf,epsfig}
\usepackage{bm}
\usepackage{mathrsfs}
\usepackage[font={small,it}]{caption}
\usepackage{bbold}
\usepackage{multirow}
\usepackage{bbm}
\usepackage{chngpage}
\usepackage{adjustbox}
\usepackage{nicefrac}
\usepackage{amssymb}
\usepackage{bigints}
\usepackage{algorithm}
\usepackage{algpseudocode}
\usepackage{natbib}
\usepackage{xr}
\RequirePackage[colorlinks,citecolor=blue,urlcolor=blue]{hyperref}

\newcommand\red[1]{{\color{black}#1}}
\newcommand{\Real}{\mathbb{R}}
\newtheorem{theorem}{Theorem}[section]

\newtheorem{proposition}[theorem]{Proposition}
\newtheorem{corollary}[theorem]{Corollary}

\newtheorem{remark}[theorem]{Remark}

\usepackage{sectsty}
\sectionfont{\fontsize{14}{15}\selectfont}

\begin{document}
\begin{center}

\textbf{\Large Informative Goodness-of-Fit  for Multivariate Distributions}
Sara Algeri$^{1}$\\
\vspace{0.5cm}
{\small $^1$ School of Statistics, University of Minnesota, \\0461 Church St SE, Minneapolis, MN 55455, USA.}\\
{\small Email: salgeri@umn.edu}\\
\end{center}

\begin{abstract}
This article introduces  an informative goodness-of-fit (iGOF) approach to study  multivariate distributions. When the null model is rejected, iGOF allows us to identify  the underlying sources of mismodeling and naturally equips practitioners with additional insights on the nature of the deviations from the true distribution.
 The  informative character of the procedure is achieved by exploiting  smooth tests and random field theory to facilitate the analysis of multivariate data. Simulation studies show that iGOF enjoys high power for different types of alternatives. The methods presented here directly address the problem of background mismodeling arising in physics and astronomy. It is in these areas that the motivation of this work is rooted. 
\end{abstract}

{\small
Multivariate goodness-of-fit,  Rosenblatt transform,  smooth tests, background mismodeling.
}
\section{Introduction}
\label{intro}

\textbf{\emph{Scientific motivations.}} 
When searching for the signals of new particles, or when aiming to detect new astronomical objects, a common difficulty arising in the analysis of the data collected by the detectors is the impossibility of correctly specifying the background distribution. In physics and astronomy, we typically refer with ``background'' or ``noise'' to the signal of all the astrophysical sources which  are not those  we aim to discover. 
 Unfortunately, since many sources contribute to the background, its distribution  is particularly difficult  to model \citep[e.g.,][]{priel, dauncey, algeriBANFF}.

 Moreover,   if the model postulated by the scientists  is rejected, it is often difficult to identify the invalidating causes.  \red{For instance, instrumental errors may lead to unexpected perturbations in the data distribution,  or there may be unpredicted cosmic sources with non-negligible contributions.  Moreover,  given the complexity of the  models investigated through physics experiments, it is often convenient to consider simplified versions of them (typically Gaussian approximations, e.g., \citet{GAMBIT1}). Hence, it is particularly important to assess the reliability of the simplified models for the data available and, if needed, provide adequate adjustments for them.}
 
 
\textbf{\emph{Statistical formulation of the problem.}} In statistical terms, these difficulties translate into two main questions arising in the statistical analysis of multivariate data. Specifically, given a random vector $\bm{X}=(X_1,\dots,X_p)$, we may wonder: 
\begin{itemize}
\item[Q1.] \emph{is the distribution of $\bm{X}$ correctly specified and, if not, in what way \red{does the true data distribution diverge} from that
hypothesized under the null hypothesis?} 
\item[Q2.] \emph{How can we improve our postulated model?} \emph{Or in other words, can we  provide a data-driven correction  \red{for} it?} 
\end{itemize}

As noted by \citet{pearson1938}, smooth tests, originally introduced by \citet{neyman37}, naturally allow us to capture and model the departure of $f$ from $g$ and thus, they offer  the framework to directly address  Q1 and Q2. 

In order to provide a high level overview on smooth tests, let $f$ be the true (unknown) probability density function (pdf) of a random variable $X\in\Real$,   $g$ is   the hypothesized density   and $G$   the respective cumulative \red{distribution} function (cdf). For example, in the above-mentioned problem of background mismodeling,  $f$ represents the true background distribution and $g$ is the background model postulated by the scientists.  A smooth model for the true probability law $f$ can be specified as 
\red{
\begin{equation}
\label{skewG}
 f(x)= g(x)d(x)=g(x)\biggl\{1+\sum_{j\geq 1}\theta_j T_j\bigl[G(x)\bigl]\biggl\},
\end{equation}
 where $d(x)=\frac{f(x)}{g(x)}$} is the likelihood ratio and  the term in the curly brackets is an orthonormal expansion for it. 
A smooth test \citep[e.g.,][]{neyman37,barton53, ledwina}   consists of testing   if any of the coefficients $\theta_j$ in \eqref{skewG} is different from  zero.  Finally, by estimating \red{$d(x)$} and constructing adequate confidence bands, it is possible to visualize the nature of the departure of $f$ from $g$.   

Despite their usefulness, smooth tests are mainly limited to the  univariate setting.   In light of this, the main methodological task of this work is to extend this framework to allow for the analysis of multivariate data.

\textbf{\emph{Main results  and organization.}}  
The   theoretical framework is presented in Section \ref{modelling}. There, we define a suitable expansion of the likelihood ratio through orthonormal functions on the unit cube.
 As shown in Sections \ref{estimation} and \ref{inference}, such representation substantially simplifies the subsequent stages of estimation, model selection and (post-selection) inference. 
In Section \ref{diagnostics}, we discuss a simple ANOVA-like testing strategy  to identify possible sources of mismodeling. Power studies are conducted via simulations in both Sections \ref{inference} and \ref{diagnostics}. 
As noted above, this work finds its main motivations in the context of astrophysical searches. Therefore, in Section \ref{bkg} we illustrate how iGOF can be used to address the problem of  mismodeling of the cosmic background considering  a realistic simulation from the    Fermi Large Area Telescope \citep{atwood}.  \red{While} this article mainly focuses on the analysis of continuous data, extensions to the discrete setting are discussed in Section \ref{discrete}.
Section \ref{conclusions} collects a summary of the results and a discussion of the limitations of  iGOF.  Technical proofs and codes are provided in the Supplementary Material.
A summary of the main notation used throughout the paper is available in the Appendix.

\section{Theoretical framework }
\label{modelling}

\subsection{\red{An orthonormal expansion for the  likelihood ratio}}
\label{Rosenblatt}
Suppose $F$ is  the true distribution function of a random vector $\textbf{X}\in \mathcal{X}\subseteq \Real^p$ and denote with $G$   its hypothesized distribution.  $F$ and $G$  are assumed to be    continuous   with densities $f$ and $g$.  Furthermore,  assume that $f(\bm{x})=0$  whenever $g(\bm{x})=0$. For every $\bm{x}=(x_1,\dots,x_p)\in  \mathcal{X}$, the hypothesized density $g$ is such that 
\[g(\bm{x})=\prod_{d=1}^p g_d(x_d|{\bm{x}}_{<d}),\]
where  $\bm{x}_{<d}=(x_1,\dots,x_{d-1})$ and  $g_1,\dots,g_p$ are suitable densities with associated cdfs and quantile functions $G_d$ and \red{$G^{-1}_{d}$}, for all $d=1,\dots,p$. The likelihood ratio between $F$ and $G$   can be specified as
\red{\begin{equation}
\label{jcd}
d(\bm{u})=\frac{f\bigl(\bm{G}^{-1}_R(\bm{u})\bigl)}{g\bigl(\bm{G}^{-1}_R(\bm{u})\bigl)},\quad\bm{u}\in[0,1]^p
\end{equation}
where $\bm{u}=(u_1,\dots,u_p)=\bigl(G_1(x_1),\dots,G_p(x_p|{\bm{x}}_{<p})\bigl)=\bm{G}_R(\bm{x})$ is the   Rosenblatt transformation  \citep{rosenblatt}\footnote{Notice that, in general, $\bm{G}_R(\bm{x}) \not\equiv G(\bm{x})$ as the Rosenblatt's transform $\bm{G}_R(\bm{x})\in[0,1]^d$ whereas the cdf $G(\bm{x})\in [0,1]$.}, and
$\bm{x}=(x_1,\dots, x_d)=\bigl(G^{-1}_1(u_1),\dots,G^{-1}_p(u_p|{\bm{x}}_{<p})\bigl)=\bm{G}^{-1}_R(\bm{u})$. }

 In the bivariate setting, for instance, let $G_1\equiv G_{X_1}$ and $G_2\equiv G_{X_2|X_1}$, i.e., the hypothesized marginal cdf of  $X_1$ and the hypothesized conditional cdf of $X_2|X_1$, respectively. Hence, \eqref{jcd} specifies as
\[d(u_1,u_2)=\frac{f_{X_1X_2}\bigl(G^{-1}_{1}(u_1),G^{-1}_{2}(u_2|x_1)\bigl)}{g_{X_1X_2}\bigl(G^{-1}_{1}(u_1),G^{-1}_{2}(u_2|x_1)\bigl)}.\]
\begin{remark}
\label{independence}
As a plausible alternative to Rosenblatt's transform, one could  choose each $G_d\equiv G_{X_d}$, which corresponds to assuming independence among the components of $\bm{X}$. In this setting, \red{if the marginal distributions are correctly specified,}  \eqref{jcd}  is  the copula density \citep[e.g.,][]{nelsen} of $\bm{X}$ under $G$. 
\red{While} this choice could simplify substantially the computations, \red{it would not allow us to test models $G$ which assume a specific dependence structure and the interest is in assessing if the joint distribution $G$ is misspecified.} Moreover, it is worth pointing out that there are situations where such transformation cannot be specified (e.g., Section \ref{bkg}). 
\end{remark}

To provide a sufficiently detailed representation of the substructures characterizing the distribution of $\bm{X}$ (see Q1   in Section \ref{intro}), a natural approach is that of expressing \eqref{jcd} by means of a suitable orthonormal  basis  in $L^2[0,1]^p$. 
For instance, let $T_{j_d}(u_d)$ be the \red{$j_d$-th} normalized shifted Legendre polynomial evaluated at $u_d=G_d(x_d|\bm{x}_{<d})$, with  $T_0(u_d)=1$, $T_1(u_d)=\sqrt{12}(u_d-0.5)$, etc. (e.g., Section 2, Supplementary Material). Each $\{T_{j_d}(u_d)\}_{j_d\geq 0}$ forms a basis in  $L^2[0,1]$. Hence, we can exploit a well known result in Hilbert space theory \citep[e.g., Proposition 2][p.50]{reedbook} which asserts that given two orthonormal bases $\{\psi_{j}\}$, $\{\phi_{k}\}$  for the Hilbert spaces $\mathcal{H}_1$,  $\mathcal{H}_2$, then  $\{\psi_{j}\otimes \phi_{k}\}$ is an orthonormal basis for $\mathcal{H}_1\otimes\mathcal{H}_2$. It follows that the tensor product basis  $\{T_{j_1,\dots,j_p}(\bm{u})\}_{j_1\dots j_p\geq 0}$ of functions
\begin{equation}
\label{Sjs}
T_{j_1\dots j_p}(\bm{u})=\prod_{d=1}^p T_{j_d}(u_d)
\end{equation}
forms an orthonormal basis on  $L^2[0,1]^p$,  the Hilbert space of square integrable function over the $p$-dimensional unit cube.

Notice that \red{while} any orthonormal basis in $[0,1]$ could be used to construct a tensor product basis in $[0,1]^p$, here we focus on the normalized shifted Legendre polynomials. This choice is justified by the fact that the latter are special cases of the so called LP-score functions \citep[e.g.,][]{LPksamples}. As discussed in Section \ref{discrete}, the latter allow for extensions to the discrete  setting.

Finally, under the assumption \red{that} $d(\bm{u})\in L^2[0,1]^p$, we can write
\begin{equation}
\label{jcd_rep}
d(\bm{u})=\sum_{j_1\geq 0,\dots,j_p\geq 0}\theta_{j_1\dots j_p}T_{j_1\dots j_p}(\bm{u}),\qquad \text{ $\bm{u}\in [0,1]^p$}
\end{equation}
with $\theta_{j_1\dots j_p}=\int_{[0,1]^p}T_{j_1\dots j_p}(\bm{u})d(\bm{u})\text{d}\bm{u}$. The expansion in \eqref{jcd_rep} follows from Theorem II.6 in \citet[][]{reedbook} and it is equivalent to say that the sum on the right-hand side converges to $d(\bm{u})$ in $L^2[0,1]^p$.

\red{
As noted by an anonymous referee,  the likelihood ratio can also be expanded on the original domain $\mathcal{X}\subseteq \Real^p$ by means of any set of bounded functions which are orthogonal with respect to $G$. In our context, this is achieved by combining the Legendre polynomials and Rosemblatt's transform. The latter provides the additional advantage of allowing us to work on the compact compact domain $[0,1]^p$. As it will become clear in Section 4, this is particularly useful as one can exploit results from random field theory to construct simultaneous confidence bands. Moreover, for visualization purposes, it may be particularly advantageous to work in the quantile domain when testing long tailed distributions. In this setting,  we may expect that only a few observations have been detected over large regions of the $\bm{X}$ domain and thus the quantile representation  allows us to magnify the differences observed over the the most ``data-abundant'' regions. An more detailed discussion of this aspect, and adequate graphical comparisons  can be found in \citet{algeri20}. 
Finally, it is worth pointing out that,  the quantile functions $G^{-1}_d$  in $\bm{G}_R^{-1}(\bm{u})$ are used in \eqref{jcd} with the only purpose of highlighting the dependence of $d$  on $\bm{u}$, when working in the quantile domain. In practice, however, estimation and inference focus entirely on \eqref{jcd_rep} (see Sections \ref{estimation} and \ref{inference}) and thus one needs not to compute $\bm{G}_R^{-1}(\bm{u})$. }

\section{Estimation}
\label{estimation}
The summations in   \eqref{jcd_rep} are taken up to infinity. However,  to make the expansion operational, it is necessary to truncate the series in \eqref{jcd_rep}  at integers values $m_1,\dots,m_p$. That is because, effectively,  the coefficients $\theta_{j_1\dots j_p}$ need to be estimated and, consequently, the more terms are  included in \eqref{jcd_rep}, the larger the variance of the resulting estimator of $d(\bm{u})$ (see Section \ref{selection} for a more detailed discussion on model selection). 

For the sake of simplifying  the notation in this section and those to follow, denote with $\mathcal{K}$ the set 
\begin{equation}
\label{kappa}
\mathcal{K}:=\biggl\{\{j_1\dots j_p\},\text{ with }  j_d=0,\dots,m_d, \text{ for all } d=1,\dots,p, \text{ and } \sum_{d=1}^p j_d\neq 0 \biggl\}
\end{equation}
of cardinality $|\mathcal{K}|=M=\prod_{d=1}^p(m_d+1)-1$. That is,   $\mathcal{K}$ contains all the  $p-$tuples $\{j_1\dots j_p\}$ of  indexes $j_d=0,\dots,m_d$,  $d=1,\dots,p$ apart from  the $p-$tuple $\{0\dots 0\}$\red{, since $\theta_{0\dots 0}=1$ (see (S.11)  in the Supplementary Material)}. 
Let $\bm{\theta}$  be the  $M\times 1$ vector of components ${\theta}_{k}$, with $k\in \mathcal{K}$. 
Similarly, denote with $\bm{T}(\bm{u})$  the $M\times 1$ vector of elements $T_{k}(\bm{u})$, $k\in \mathcal{K}$.  

Consider $\bm{x}_1,\dots,\bm{x}_n$,  a sample of $n$ i.i.d. observations from  $\bm{X}$, and let $\bm{U}=\bm{G}_R(\bm{X})$ be the respective Rosenblatt transformation.  
Denote with $\bm{u}_1,\dots,\bm{u}_n$ the sample of elements $\bm{u}_i=\bm{G}_R(\bm{x}_i).$
The parameter ${\bm{\theta}}$ can be estimated by means of  the vector $\widehat{\bm{\theta}}$ of components
\red{
\begin{equation}
\label{theta_est}
\widehat{\theta}_{k}=\frac{1}{n}\sum_{i=1}^n T_{k}\bigl(\bm{G}_R(\bm{x}_i)\bigl)=\frac{1}{n}\sum_{i=1}^n T_{k}(\bm{u}_i)\quad\text{ for all $k\in \mathcal{K}$ },
\end{equation}}
The mean and covariance matrix of  $\widehat{\bm{\theta}}$ and an estimator of  $d(\bm{u})$  are given in Proposition \ref{moments_prop}. 
\begin{proposition}
\label{moments_prop}
The likelihood ratio $d\bigl(\bm{u}\bigl)$  is the density of the random vector $\bm{U}$ and
\begin{equation}
\label{moments}
E[\widehat{\bm{\theta}}]=\bm{\theta}\quad\text{and}\quad \text{Cov}(\widehat{\bm{\theta}})=\bm{\Sigma}
\end{equation}
where $\bm{\Sigma}$ has diagonal elements $\frac{\sigma^2_{k}}{n}=\frac{1}{n}V\bigl[T_{k}(\bm{U})\bigl]$ and non-diagonal elements
$\frac{\sigma_{k,h}}{n}=\frac{1}{n}\text{Cov}\bigl[T_{k}(\bm{U}),T_{h}(\bm{U})\bigl]$, with $k,h\in \mathcal{K}$.
 Furthermore, if $F\equiv G$, the equalities in \eqref{moments} reduce to
\begin{equation}
\label{moments0}
E[\widehat{\bm{\theta}}]=\bm{0} \quad\text{and}\quad \text{Cov}(\widehat{\bm{\theta}})=\frac{1}{n}\bm{I}_M,
\end{equation}
where $\bm{0}$ is the $M\times1$ zero vector and $\bm{I}_M$ is the $M\times M$ identity matrix.

Finally, an estimator of $d(\bm{u})$ is
\begin{align}
\label{jcd_est}
\widehat{d}(\bm{u})&=1+\widehat{\bm{\theta}}'\bm{T}(\bm{u}),
\end{align}
and has variance $V\Bigl[\widehat{d}(\bm{u})\Bigl]=\bm{T}(\bm{u})'\bm{\Sigma}\bm{T}(\bm{u})$.
\end{proposition}

Combining \eqref{skewG}, \eqref{jcd} and \eqref{jcd_est} an estimate of $f$ is 
\begin{align}
\label{skewG_est}
\widehat{f}(\bm{x})=g(\bm{x})\widehat{d}(\bm{x})=g(\bm{x})\bigl[1+\widehat{\bm{\theta}}'\bm{T}\bigl( \bm{G}_R(\bm{x})\bigl)\bigl]
\end{align}
Notice that the estimator $\widehat{f}$   incorporates the information carried by the hypothesized model $g$; whereas, the estimator in the square brackets provides a data-driven correction for it. 
Furthermore, define the integrated squared bias (ISB) of $\widehat{d}(\bm{u})$ to be
\begin{equation}
\label{ISBdef}
ISB=\int_{[0,1]^p}\Bigl(E[\widehat{d}(\bm{u})]-d(\bm{u})\Bigl)^2\text{d}\bm{u}.
\end{equation}
\begin{figure*}[!h]
\begin{tabular*}{\textwidth}{@{\extracolsep{\fill}}@{}c@{}c@{}}
\includegraphics[width=70mm ]{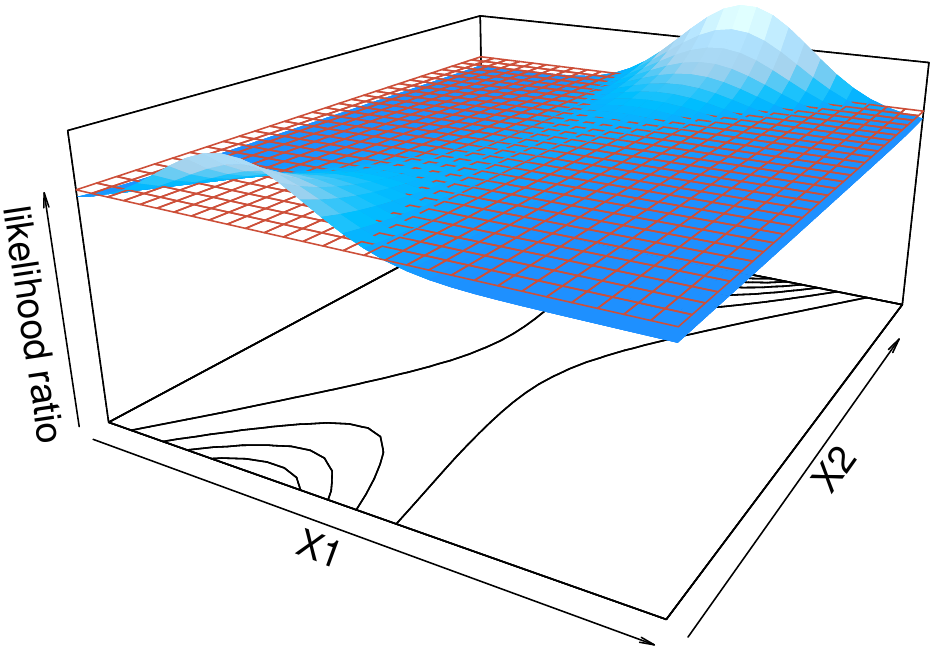} & \hspace{-1cm} \includegraphics[width=71mm ]{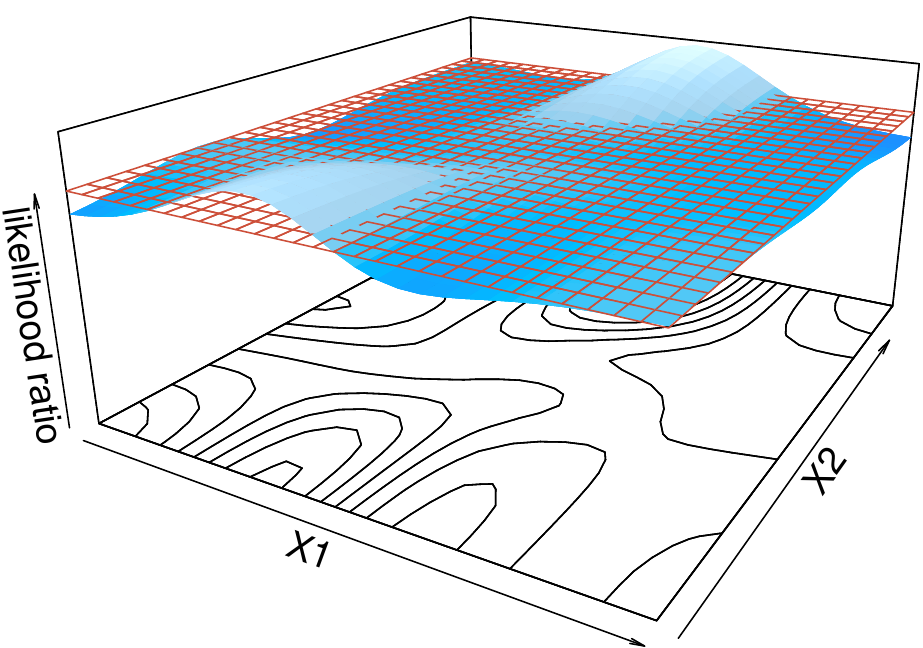}\\
\end{tabular*}
\caption{True (left panel) and estimated  (right panel) likelihood ratio for Example I.  The  estimate on the right has been obtained via \eqref{jcd_est} with $m_1=4$ and $m_2=3$. The components of the $\widehat{\bm{\theta}}$ vector have been selected via the AIC criterion in \eqref{BIC}. 
}
\label{fig1}
\end{figure*}
From Proposition \ref{ISB_prop} it follows that the closer $g$ is to $f$ in terms of squared normalized distance the lower the ISB of $\widehat{d}(\bm{u})$. 
\begin{proposition}
\label{ISB_prop}
The integrated squared bias of the estimator in \eqref{jcd_est} is
\red{
\begin{equation}
\label{ISB}
ISB=\bigintsss_{[0,1]^p}\biggl(\frac{f\bigl(\bm{G}^{-1}_R(\bm{u})\bigl)- g\bigl(\bm{G}^{-1}_R(\bm{u})\bigl)}{g\bigl(\bm{G}^{-1}_R(\bm{u})\bigl)}\biggl)^2\text{d} \bm{u}-\bm{\theta}'\bm{1}=\bigintsss_{\mathcal{X}}\biggl(\frac{f(\bm{x})- g(\bm{x})}{g(\bm{x})}\biggl)^2g(\bm{x})\text{d} \bm{x}-\bm{\theta}'\bm{1}
\end{equation}}
where $\bm{1}$ is the  $M\times 1$ unit vector.
\end{proposition}
The estimate in \eqref{skewG_est} is essentially that of a smooth model \citep[e.g.,][]{rayner90}, that is, a smoothed version of the true underlying probability function. Similarly to the smooth model proposed by \citet{barton53} in the univariate setting, the estimator in \eqref{skewG_est} may lead to estimate that are not \emph{bona-fide}, i.e, they may be negative and/or they may not integrate/sum up to one. In this manuscript we focus on \eqref{skewG_est} mostly for the sake of mathematical convenience \red{when constructing simultaneous confidence bands in Section \ref{inference}}. Nonetheless, bona-fide estimators can be constructed similarly to the univariate case as described  in \citet{algeri_zhang}.

\emph{\textbf{Example I.}} In direct searches for dark matter, the dominant  background sources are neutron recoils which may produce signals mimicking those expected from dark matter candidates \citep[e.g.,][]{westerdale}. As a toy example, suppose we are interested in assessing the validity of a given distribution for the nuclear recoil background specified over the  energy region  $\mathcal{X}=[5,20]KeVnr\times[0,17]KeVnr$. Each observations in $\mathcal{X}$ corresponds  to the scintillation of photons ($X_1$) and  ionization electrons ($X_2$)  \citep[e.g.,][]{aprile}.  The hypothesized background distribution, $G_{X_1X_1}$, is that of a truncated bivariate normal with mean vector $(12,8)$, variances $8$ and $12$ and covariance $2$. Moreover,  suppose that one additional  background source  is present. The latter is also a bivariate normal with the same mean vector, variances $4$ and $20$ and covariance $5$. Thus, the true model, $F_{X_1X_2}$, involves a mixture of two, overlapping truncated  bivariate Gaussians with mixture parameter $0.15$. In order to estimate the likelihood ratio, set $G_1=G_{X_1}(x_1)$ and $G_2=G_{X_2|X_1}$. The estimated likelihood ratio, obtained over a sample of $n=5,000$, is shown in the   right panel of Figure \ref{fig1}, whereas the   left panel shows the true likelihood ratio.  A closed form expression for the estimate shown on the right panel is given in equation (S.34) in the Supplementary Material.
\red{The   estimate obtained    recovers the main departures from uniformity. Specifically, the second mixture component contributing to $F_{X_1X_2}$  leads  the an inflation of the variance of $X_2$;  as a result, $F_{X_1X_2}$ exhibits  higher tails than $G_{X_1X_2}$ in the direction of the first eigenvector. Such departure becomes more and more prominent when moving from the center of the distribution towards the truncation points $X_2=0$ and $X_2=17$. Finally, the contours of $\widehat{d}(\bm{u})$  } highlight that the estimator is rather noisy. Therefore, it is  important to investigate the properties of   \eqref{jcd_est} to  assess the significance of the deviations observed.

\section{Inference and model selection}
\label{inference}
\subsection{Pre-selection inference}
A smooth test for  $H_0:G\equiv F$ versus $H_1:G\not\equiv F$ consists in reformulating the problem as a test for uniformity of $\bm{U}$. 
Specifically, \eqref{jcd} implies that   $F\equiv G$ whenever $d(\bm{u})=1$, and thus  
\begin{equation}
\label{test0}
\begin{split}
H_0: d(\bm{u})=1 \quad\text{for all $\bm{u}\in [0,1]^p$} \quad\text{versus} \quad H_1 : \exists \bm{u}\in [0,1]^p \text{ such that } d(\bm{u})\neq 1.
\end{split}
\end{equation}
It is easy to see that $d(\bm{u})=1$ for all $\bm{u}\in [0,1]^p$, when  all  ${\theta}_{k}$, $k\in\mathcal{K}$, are identically equal to zero. Hence,  in practice, we test
\begin{equation}
\label{test}
H_0: \bm{\theta}= \bm{0}  \qquad\text{vs}\qquad H_1 : \bm{\theta}\neq \bm{0}.
\end{equation}
Notice that $H_0$ in \eqref{test0} implies $H_0$ in \eqref{test}, but the opposite is not true in general. Whereas, $H_1$ in \eqref{test}  does imply  $H_1$ in \eqref{test0}. 
With a little
abuse of nomenclature, in this section and those to follow, we will refer to $G$ as the ``null model''. Furthermore, we will refer to  $H_0$ in \eqref{test0} when generically saying  ``under $ H_0$''. However, most of the results presented here, only require validity of the  milder  $H_0$ in \eqref{test}.

To conduct our inference, we consider the so-called \emph{deviance} test statistics, i.e.,
\begin{equation}
\label{deviance}
D=n\widehat{\bm{\theta}}'\widehat{\bm{\theta}}.
\end{equation} 
Its asymptotic null distribution   is given in Theorem \ref{normality}.
\begin{theorem}
\label{normality}
If $H_0$ is true,  then 
\vspace{-0.2cm}
\begin{equation}
\label{thetaH0}
\sqrt{n}\widehat{\bm{\theta}}\xrightarrow d N(\bm{0},\bm{I}), \quad\text{as $n\rightarrow\infty$}
\end{equation}
\vspace{-0.2cm}
where $N(\bm{0},\bm{I})$ denotes a standard multivariate normal distribution. Furthermore,
\begin{equation}
\label{DH0}
D \xrightarrow d \chi^2_{M}, \quad\text{as $n\rightarrow\infty$},
\end{equation}
where $M$ is the size of $\widehat{\bm{\theta}}$. 
\end{theorem}
\red{The asymptotic distribution of the random field $\widehat{d}(\bm{u})$ is derived in Theorem \ref{dhat_cor} below. This result is particularly useful for us to construct simultaneous confidence bands as described in Section \ref{selection}.}
\begin{theorem}
\label{dhat_cor}
Denote with $\{\widehat{d}(\bm{u})\}$ the random field indexed by $\bm{u}\in [0,1]^p$ with components as in \eqref{jcd_est}.  Moreover, assume that $\widehat{\theta}_k=o(n^{-1/2})$ for all $k\not\in \mathcal{K}$.  If  $H_0$ is true, 
\begin{equation}
\label{rf}
\Biggl\{\frac{\widehat{d}(\bm{u})-1}{\sqrt{\frac{1}{n}\bm{T}(\bm{u})'\bm{T}(\bm{u})}}\Biggl\}\xrightarrow d \bm{Z}(\bm{u}), \quad\text{as $n\rightarrow\infty$,}
\end{equation}
where $\bm{Z}(\bm{u})$ denotes a Gaussian random field with mean zero, unit variance and covariance function 
$\text{Cov}\Bigl( \bm{Z}(\bm{u}), \bm{Z}(\bm{u}^\dag)\Bigl)=\frac{\bm{T}(\bm{u})'\bm{T}(\bm{u}^\dag)}{\sqrt{\bm{T}(\bm{u})'\bm{T}(\bm{u})\bm{T}(\bm{u}^\dag)'\bm{T}(\bm{u}^\dag)}}.$
\end{theorem}

At this stage, constructing inference on the basis of Theorems \ref{normality} and  \ref{dhat_cor} would be  tempting. However, to guarantee the validity of our results we must take into account that, when estimating the likelihood ratio in \eqref{jcd_est}, a model selection procedure is  likely to be  implemented. Unfortunately, when a model is selected by a pool of possibilities, such process introduces an additional source of variability and thus the resulting inference is automatically affected \citep[e.g.,][]{berk}. Section \ref{selection}   addresses this aspect directly.

\subsection{Post-selection inference}
\label{selection}

The estimate of the likelihood ratio considered so far involves up to $M$   functions $T_{k}(\bm{u})$. Nonetheless, it is possible that not all of these $M$ terms are needed to capture the departures of $G$ from $F$ and indeed, it is often convenient to remove some of them  to avoid unnecessary sources of noise. Various criteria have been proposed in literature for density estimation and smooth models \citep[e.g.,][]{LPmode, algeri20} and which can be easily extended to the multivariate setting. Here, we focus on the approach of  \citet{LPmode} and which specifies as follows.

Let $\widehat{\theta}_{(k)}$ be the \red{$k$-th} largest $\widehat{\theta}_{k}$ estimate in order of magnitude, \red{for $k\in \mathcal{K}$}, i.e.,  $\widehat{\theta}_{(1)}^2\geq \widehat{\theta}_{(2)}^2\geq\dots\geq \widehat{\theta}_{(M)}^2$. Select the  $K$ largest coefficients which maximize either
\begin{equation}
\label{BIC}
\text{BIC}(K) = \sum_{(k)=1}^K\widehat{\theta}^2_{(k)} - \frac{K \log n }{n}\quad\text{or}\quad  \text{AIC}(K) = \sum_{(k)=1}^K\widehat{\theta}^2_{(k)} - \frac{2K}{n}.
\end{equation}
Notice that, as defined in \eqref{kappa}, each $k\in\mathcal{K}$ is a $p-$tuple of indexes $j_1\dots j_p$, whereas $(k)$ is the integer value corresponding to the order of magnitude of the respective coefficient $\widehat{\theta}_{k}$. Hence, the summations in \eqref{BIC} and those to follow are taken over $(k)=1,\dots,K$, that is, the $K$ $p-$tuples of indexes $j_1\dots j_p$ with the \red{$K$-th} largest estimates $\widehat{\theta}_{k}$. 

An estimate of $d(\bm{u})$, is then selected via \eqref{BIC} from the family of estimators
\begin{equation}
\label{estAIC}
\widehat{d}_{(K)}(\bm{u})=1+\sum_{(k)=1}^{K}\widehat{\theta}_{(k)}T_{(k)}(\bm{u}),\quad\text{$K=1,\dots,M$}
\end{equation}
where the subscript $(K)$ on the left-hand-side is used to emphasize that the estimator in \eqref{estAIC} includes only the \red{$K$-th} largest $\widehat{\theta}_k$ estimated coefficients. 
Clearly, the choice of BIC or AIC is arbitrary and, from a practical standpoint, \red{when $n>7(\approx e^2)$, the BIC assigns a heavier penalty than AIC for increasing values of $K$ and, consequently, it often leads to smoother estimators than AIC.}

The selection rules in \eqref{BIC} compare $M$ possible models assuming that each $m_d$, for $d=1,\dots,D$ was fixed before the researcher looked at the data. Valid post-selection inference can then be constructed as in \red{Corollaries \ref{naive} and \ref{naive2}}. The respective proofs are provided in the Supplementary material.

\begin{corollary}
\label{naive}
Denote with  $\widehat{d}_{(K*)}$  the estimator of $d(\bm{u})$ selected via \eqref{BIC}, and let $D_{(K^*)}=\sum_{(k)=1}^{K^*}\widehat{\theta}^2_{(k)}$ be the respective deviance statistics.
As  $n\rightarrow \infty$, a  valid post-selection bound for the p-value to test \eqref{test0} is  
\begin{equation}
\label{pvalue}
\text{p-value}_{adj}=P(\chi^2_{M }>D_\text{obs}),
\end{equation}
where $D_\text{obs}$ is the value of $D_{(K^*)}$ observed.
\end{corollary}
Where the bound in \eqref{pvalue}, follows from the fact that the estimators in \eqref{estAIC} are nested, for all $K=1,\dots,M-1$,  and thus each  \red{$D_{(K)}=\sum_{(k)=1}^K\widehat{\theta}^2_{(k)}$ } is stochastically lower or equal than $D_{(M)}=\sum_{(k)=1}^M\widehat{\theta}^2_{(k)}$, that is,  for all $K=1,\dots,M-1$, $P(D_{(K)}>D_\text{obs})$ is smaller than $P(D_{(M)}>D_\text{obs})$.

In order to grasp further insights on the deviations of $G$ from $F$,   it is worth constructing adequate confidence bands. This can be done, while accounting for post-selection adjustments, as in Corollary \ref{naive2}. 
\begin{corollary}
\label{naive2}
Denote with  $\widehat{d}_{(K^*)}$  the estimator of $d(\bm{u})$ selected via \eqref{BIC}, and let
 $SE_0\bigl[\widehat{d}_{(K^*)}(\bm{u})\bigl]$ be  its  standard error   under $H_0$. \red{Moreover, assume that,
\begin{equation}
\label{stoc}
\Biggl[\sup_{\bm{u}}\biggl\{\frac{\widehat{d}_{(K)}(\bm{u})-1}{SE_0\bigl(\widehat{d}_{(K)}(\bm{u})\bigl)}\biggl\} \biggl| K^*=K\Biggl]\preceq \Biggl[\sup_{\bm{u}}\biggl\{\frac{\widehat{d}_{(M)}(\bm{u})-1}{SE_0\bigl(\widehat{d}_{(M)}(\bm{u})\bigl)}\biggl\} \biggl| K^*=K\Biggl]
\end{equation}
for all   $K=1,\dots,M-1$.} Valid (post-selection adjusted) \red{simultaneous} $(1-\alpha)\%$ confidence regions, under $H_0$ in \eqref{test0}, are 
\begin{equation}
\label{CIband}
\biggl[1-c_{\alpha/2} SE_0\bigl[\widehat{d}_{(K^*)}(\bm{u})\bigl],1+c_{\alpha/2}SE_0\bigl[\widehat{d}_{(K^*)}(\bm{u})\bigl]\biggl]\qquad \text{for all $\bm{u}\in [0,1]^p$}
\end{equation}
with $c_{\alpha/2}$ such that 
\begin{equation}
\label{calpha}
P\Biggl(\sup_{\bm{u}}\biggl\{\frac{\widehat{d}_{(M)}(\bm{u})-1}{SE_0\bigl[\widehat{d}_{(M)}(\bm{u})\bigl]}\biggl\}>c_{\alpha/2} \Biggl|H_0\Biggl)=\frac{\alpha}{2}.
\end{equation}
\end{corollary}
\red{In \eqref{stoc},  ``$\preceq$'' indicates that the left-hand-side is stochastically lower or equal than the right-hand-side. Intuitively, the validity of \eqref{stoc} in practical settings follows from the fact that $\widehat{d}_{(M)}$  is the least smooth among all the estimators considered; thus, we expect that  the  random field resulting from $\widehat{d}_{(M)}$ has the largest probability of crossing the fixed level $c_{\alpha/2}$. 

The confidence bands in \eqref{CIband} are constructed around 1, not around $\widehat{d}_{(K^*)}(\bm{u})$. That is because 
  $\widehat{d}_{(K^*)}(\bm{u})$ only accounts for the $K^{*}$ largest terms in \eqref{jcd_rep}, therefore, it is
 a biased estimator of $d(\bm{u})$ (see Proposition \ref{ISB_prop}).
 It follows that, when the bias is large, confidence bands constructed around $\widehat{d}_{(K^*)}(\bm{u})$ would be shifted
away from the true density  $d(\bm{u})$. However, under $H_0$, both the bias at a point $\bm{u}$
and the integrated bias are equal to zero. Hence, \eqref{CIband} are reliable confidence bands under $H_0$.}

From a theoretical perspective, a highly  non-trivial aspect in the construction of \eqref{CIband} is the estimation of the quantile $c_{\alpha/2}$. Probabilities such \eqref{calpha} are known in literature as \emph{excursion probabilities} \citep[e.g.,][]{adler2000} and  cannot be expressed in closed form. A possible solution for constructing the confidence bands in \eqref{CIband}, is to proceed by estimating $SE_0\bigl[\widehat{d}_{(K^*)}(\bm{u})\bigl]$ and $c_{\alpha/2}$ via Monte Carlo simulations \citep[see][Algorithm 1]{algeri_zhang}.  Unfortunately, in the most crucial (astro)physical searches the level of significance required to claim a new discovery is typically in the order of $\alpha=10^{-7}$ \citep[e.g.,][]{lyons2015}, and thus  Monte Carlo simulations  may be computationally prohibitive. This   is further aggravated when dealing with complex models for which even a single Monte Carlo replicate can be highly expensive in terms of both computational and time resources.

As a valid alternative, for continuous  $F$ and $G$, accurate approximations for \eqref{calpha} under mild smoothness conditions exist \citep[e.g.,][]{taylor2008}. 
In our setting, smoothness follows from the fact that the random field in \eqref{rf} and the respective limit  can be written as a linear combination of the functions $\frac{T_k(\bm{u})}{\sqrt{\sum_{k\in\mathcal{K}}T^2_k(\bm{u}) }}$ 
(see proof of Theorem \ref{dhat_cor} in the Supplementary Material) which are composition of Legendre polynomials, and thus, admit infinite partial derivatives. 

An approximation for the left-hand side of \eqref{calpha}  is
\begin{equation}
\label{LK}
\bigl(1-\Phi(c_{\alpha/2})\bigl)+\mathcal{L}_1\frac{e^{-\frac{c_{\alpha/2}^2}{2}}}{\pi}+\mathcal{L}_2\frac{e^{-\frac{c_{\alpha/2}^2}{2}}}{\sqrt{2}\pi^{3/2}}+O\biggl(\exp\Bigl(-\frac{\gamma c_{\alpha/2}^2}{2}\Bigl)\biggl),\quad\text{as $n\rightarrow\infty$,}
\end{equation}
\noindent for some $\gamma>1$ \citep{takemura}. In \eqref{LK}, $\mathcal{L}_1$ and $\mathcal{L}_2$ are constant known as Lipischitz-Killing curvatures and are typically estimated numerically \citep[e.g.,][]{TOHM}. Notice that the error rate in \eqref{LK} decreases exponentially fast, as $\alpha\rightarrow \infty$. Therefore, this solution is particularly amenable to overcome the issues arising  when dealing with stringent significance requirements.

\begin{figure*}[htb]
\begin{tabular*}{\textwidth}{@{\extracolsep{\fill}}@{}c@{}c@{}}
\includegraphics[width=72mm ]{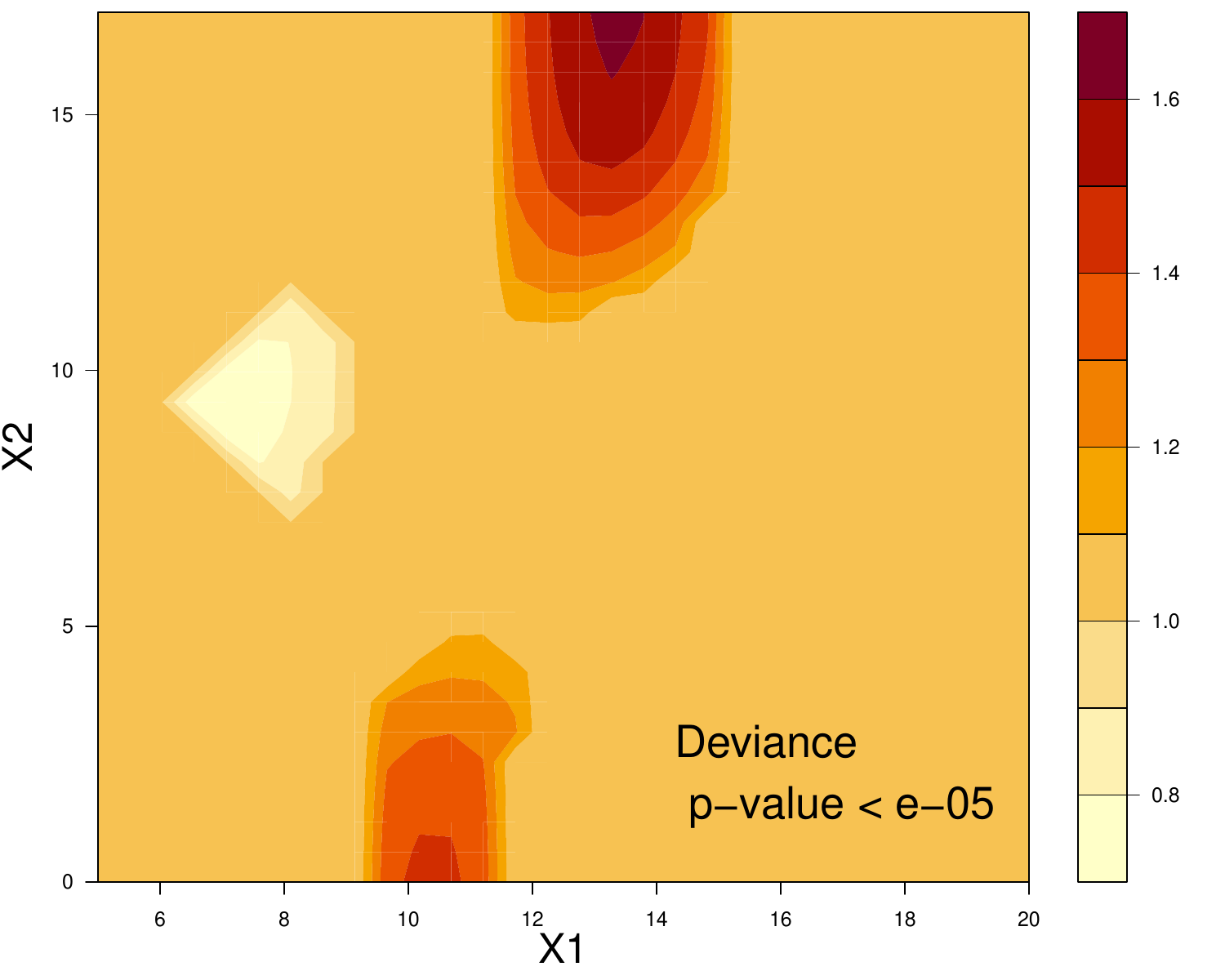}&\includegraphics[width=72mm ]{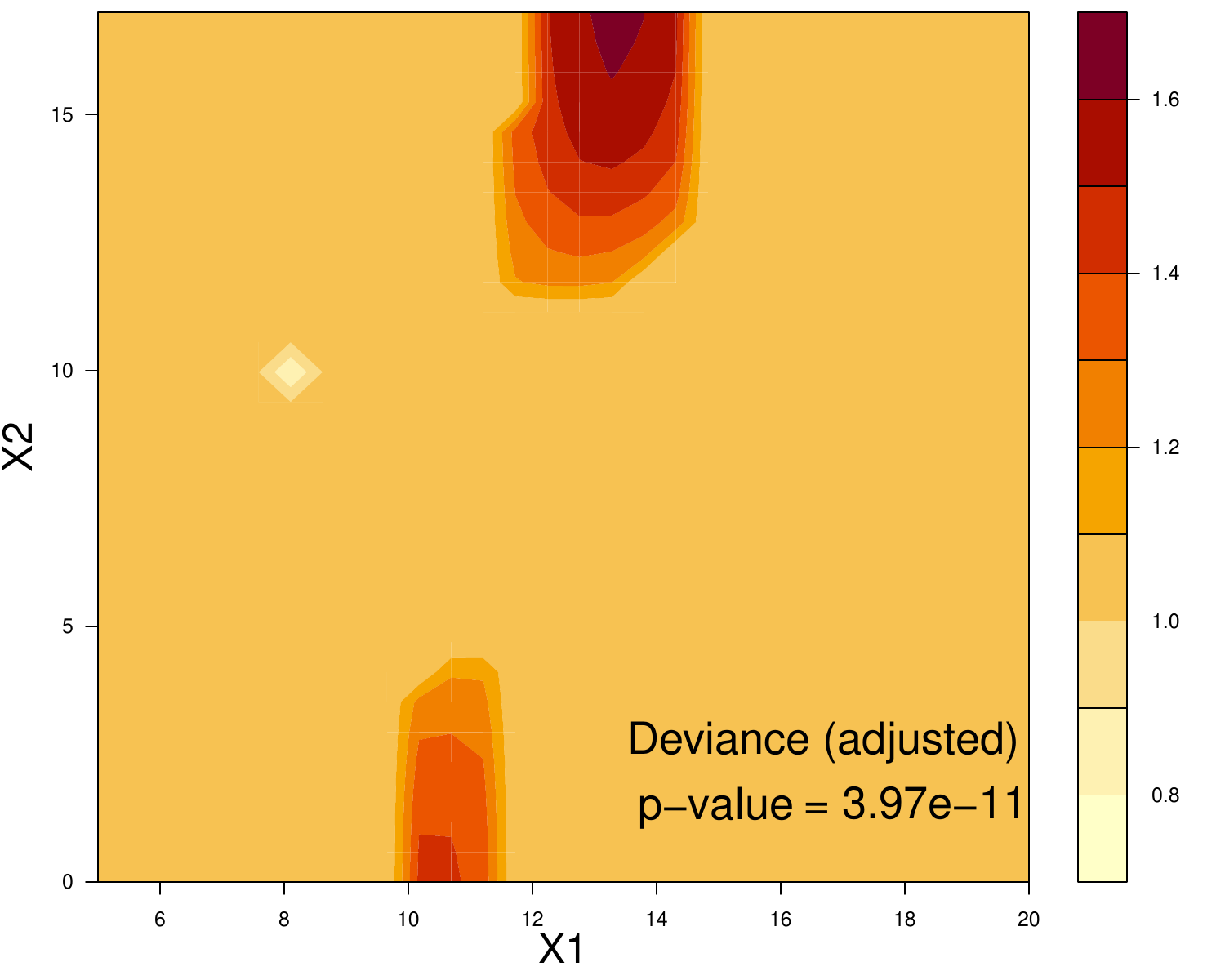} \\
\end{tabular*}
\caption{Simulated and approximated confidence regions for Example I. The left panel corresponds to the (post-selection) confidence regions and deviance p-value   obtained via a simulation of size $B=10,000$. The right panel shows to the (post-selection adjusted) confidence regions and deviance  p-value computed as in \eqref{pvalue} and \eqref{CIband}. Darker shades correspond to significant deviations of the estimated likelihood ratio above one. Lighter shades correspond to significant deviations below one. 
}
\label{fig2}
\end{figure*}
\begin{table}[!h]
\fontsize{9}{9}\selectfont{
 \centering
\begin{tabular}{|c|cccccc|}
\noalign{\global\arrayrulewidth0.05cm}
 \hline
 \noalign{\global\arrayrulewidth0.05pt}
  & $n=500$ & $n=1000$  & n=2000 & $ n=5000$ & $ n=7000$  & $ n=10,000$       \\
 \hline
Type I error &0.0540 & 0.0500  & 0.0499  &  0.0482 &  0.0508  &0.04930\\
($\pm$ SE) & ($\pm$ 0.0023) & ($\pm$ 0.0022) &($\pm$ 0.0022)   &  ($\pm$ 0.0021) &    ($\pm$ 0.0022)  & ($\pm$ 0.0022) \\
 \hline
 Power&  0.2157 &0.4456   &0.8063     &  0.9995    &1.0000& 1.0000 \\
  ($\pm$ SE)&  ($\pm$ 0.0041) &($\pm$ 0.0050)    &($\pm$ 0.0040)    &  ($\pm$ 0.0002)    &($\pm$ 0.0000)  &  ($\pm$ 0.0000)\\
\noalign{\global\arrayrulewidth0.05cm}
 \hline
 \noalign{\global\arrayrulewidth0.05pt}
\end{tabular}
\caption{Simulated probability of type I error and power for Example I considering different sample sizes. The nominal level is chosen to be $\alpha=0.05$. Each simulation involves $B=10,000$ replicates.}
\label{time}}
\end{table}
As one may expect,  the simplicity  of the post-selection adjustments  in \eqref{pvalue} and  \eqref{CIband}  comes with a price. Specifically, they can be rather conservative for increasing values of $M$. 
However, as shown below for Example I and in the sections to follow,   \eqref{pvalue} still leads to high power even if the sample size is only moderately large. Similarly, \eqref{CIband} can be quite accurate and match closely the confidence regions obtained by simulating directly the distribution of \eqref{rf}, while repeating the selection process at each replicate.

\emph{\textbf{Example I (continued).}} The estimate of the likelihood ratio in the right panel of Figure \ref{fig1} has been obtained by setting $m_1=4$ and $m_2=3$ and selecting the  terms of the respective tensor basis via the AIC rule in \eqref{BIC}. 
 The AIC procedure selects $9$ terms out of $M=19$.
 The post-selection adjusted p-value and $95\%$ confidence regions are shown in the right panel of Figure \ref{fig2}. 
The confidence contours are constructed by setting equal to one all the values of $\widehat{d}$ contained within the bands in \eqref{CIband}. Whereas the quantile $c_\alpha$  has been calculated by solving
\begin{equation}
\label{solver}
\bigl(1-\Phi(c_{\alpha/2})\bigl)+\mathcal{L}_1\frac{e^{-\frac{c_{\alpha/2}^2}{2}}}{\pi}+\mathcal{L}_2\frac{e^{-\frac{c_{\alpha/2}^2}{2}}}{\sqrt{2}\pi^{3/2}}-\frac{\alpha}{2}=0
\end{equation}
and estimating  $\mathcal{L}_1$ and $\mathcal{L}_2$ by means  of the \texttt{R} package \texttt{TOHM} \citep{TOHMpkg} as described in \citet{TOHM}. This approach led to $c_{\alpha}=3.5568$.
The confidence contours suggest that the most prominent deviations occur in correspondence of  the regions $[10,12]\times[0,5]$ and $[12,15]\times[12,17]$. Here, the estimator of $d(\bm{u})$  shows significant deviations above one and thus we conclude that the postulated model underestimates the truth over these areas. The presence of significant departures of $G_{X_1X_2}$ from $F_{X_1X_2}$ are confirmed by the deviance test (adjusted p-value $\sim 3.97 \cdot 10^{-11}$). 
The left panel of Figure   \ref{fig2} shows the confidence regions and deviance p-value obtained by means of a Monte Carlo simulation involving $B=10,000$ replicates. The selection procedure has been implemented at each replicate.  \red{While} more conservative, the confidence regions computed via \eqref{CIband} and \eqref{solver}, approximate reasonably well those obtained via simulation.

Finally, we investigate the probability of type I error and the power of the deviance test based on \eqref{pvalue}. Table \ref{time} reports the results obtained considering a suite of five simulations, each of size $B=10,000$, conducted using five different sample sizes.  For all $n$ considered, the probability of type I error observed is approximately the same than the nominal level $\alpha=0.05$. Whereas, the power increases rapidly with $n$.  For the smallest samples sizes considered, i.e., $n=500$ and $n=1000$, the power is rather low ($\sim 22\%$ and $\sim 45\%$, respectively). However, it has to be noted  that, in our example, the mixture parameter is $0.15$; therefore the deviations from the postulated model  effectively account for only $\sim 75$ and $\sim 150$ data points when $n=500$ and $n=1000$, respectively.

\section{iGOF-diagnostic analysis}
\label{diagnostics}
The constructs introduced so far allow us to assess the validity of the postulated model, obtain an estimate of the likelihood ratio test to visualize where and how departures of $g$ from $f$ occur, and construct a data driven correction for the initial model $g$ (equation \ref{skewG_est}).  Unfortunately, however, a visual inspection  is only possible when $p\leq 3$. Nevertheless, when $p>3$, more insights on the sources of mismodeling affecting $G$ can be obtained by conducting an ANOVA-like analysis where random sub-vectors of $\bm{X}$ are tested individually, from the largest to the smallest. 

Without loss of generality, let $\bm{X}_q=(X_1,\dots,X_q)$  be the random  collecting the first $q<p$ components of $\bm{X}$. Denote with $F_q$ the true cdf of $\bm{X}_q$ and let $G_q$ be its postulated cdf. Moreover, assume that the density of $G_q$   can be specified as
\begin{equation}
\label{C1}
g_{q}\bigl(\bm{x}_q\bigl)=\prod_{d=1}^q g_d\bigl(x_d|\bm{x}_{<d}\bigl)\quad\text{for all $d=1,\dots,q$.}
\end{equation}
As in \eqref{jcd} and \eqref{jcd_rep}, we can then express the likelihood ratio of $\bm{X}_q$ on the $q$ dimensional unit cube via
\begin{equation}
\label{LRq}
d(\bm{u}_q)=\sum_{j_1\geq 0,\dots,j_q\geq 0}\theta_{j_1\dots j_q}T_{j_1\dots j_q}(\bm{u}_q),\quad\bm{u_q}\in[0,1]^q
\end{equation}
where $\bm{u}_q=\bigl(G_1(x_1),\dots,G_q(x_q|\bm{x}_{<q})\bigl)$, and thus $\bm{u}_q$ is a sub-vector of $\bm{u}=\bm{G}_R(\bm{x})$.
Whereas, similarly to \eqref{Sjs}, one can write the tensor basis functions $T_{j_1\dots j_q}$ as
\begin{equation}
\label{Sj2}
T_{j_1\dots j_q}(\bm{u}_q)=\prod_{d=1}^q T_{j_d}(u_d)=\prod_{d=1}^p T_{j_d}(u_d) \quad \text{ with }  j_d=0, \text{ for  all } d=q+1.
\end{equation}
The last equality follows from the fact that $T_0\bigl(G(x_d|\bm{x}_{<d})\bigl)=1$ for all $d=1,\dots,p$, and thus each $T_{j_1\dots j_q}(\bm{u}_q)=T_{j_1\dots j_q0\dots 0}(\bm{u})$. Consequently, the $\theta_{j_1\dots j_q}$ coefficients are equal to  $\theta_{j_1\dots j_p}$ whenever $j_d=0$,   for  all $d=q+1$. 
As a result, we can easily perform inference for $\bm{X}_q$ by means of the estimators $\widehat{\theta}_k$ in \eqref{theta_est}, without the need of an entirely new estimation procedure.

Specifically, denote with $\mathcal{K}_q$ and $\mathcal{K}^*$ the subsets of $\mathcal{K}$ in \eqref{kappa}
\begin{align}
\label{kappaq}
\mathcal{K}_q&:=\biggl\{k=\{j_1\dots j_p\}\in \mathcal{K} \text{ with }  j_d=0, \text{ for  all } d=q+1,\dots, p\biggl\}\\
\label{kappaK}
\mathcal{K}^*&:=\biggl\{k=\{j_1\dots j_p\}\in \mathcal{K} \text{ with }  (k)\leq K^*\biggl\}
\end{align}
of cardinality $|\mathcal{K}_q|= M_q=\prod_{d=1}^q(m_d+1)-1$ and $|\mathcal{K}^*|=K^*$. Recall that $K^*$ is the value minimizing either the AIC or BIC in  \eqref{BIC}, and thus,
$\mathcal{K}^*$ collects all the $p-$tuple of indexes in $\mathcal{K}$ which have been ultimately selected when constructing the estimator $\widehat{d}_{(K^*)}$ and the deviance statistics $D_{(K^*)}$ in Corollaries \ref{naive} and \ref{naive2}. To test
\begin{equation}
\label{testQ}
H_0: G_q=F_q\quad\text{versus} \quad H_1: G_{q}\neq F_{q}
\end{equation}
we may consider the test statistics $D_q=n\sum_{k\in \mathcal{K}_q }\widehat{\theta}^2_k$ and proceed as in Theorem \ref{normality}.
Whereas, valid post-selection inference can be obtained as in Theorem \ref{anova_theo}.
\begin{theorem}
\label{anova_theo}
As  $n\rightarrow \infty$, a  valid post-selection bound for the p-value to test \eqref{testQ} is
\begin{equation}
\label{test3}
\text{p-value}_{q,adj}=P(\chi^2_{M_{q}}>D_{obs}),
\end{equation}
where $D_{obs}$ being the value of the test statistics
\begin{equation}
\label{Dq}
D^*_q=n\sum_{k\in \mathcal{K}_q\cap\mathcal{K}^* }\widehat{\theta}^2_k
\end{equation}
 observed,   $\mathcal{K}_q$ and $\mathcal{K}^*$ as in \eqref{kappaq} and \eqref{kappaK} and $\widehat{\theta}_k$ as in \eqref{theta_est}.
\end{theorem}
Theorem \ref{anova_theo} follows directly from \eqref{LRq} and \eqref{Sj2}, orthogonality of the $T_k$ functions, and from condition \eqref{C1}. 
\begin{center}
\begin{table}[!h]
\fontsize{9}{9}\selectfont{
 \centering
\begin{tabular}{|c|c|c|c|}
\noalign{\global\arrayrulewidth0.05cm}
 \hline
 \noalign{\global\arrayrulewidth0.05pt}
 \vspace{-0.1cm}
   &  &   &       \\
\textbf{Variable}  & \textbf{True} ($F$) &  \textbf{Hypothesized} ($G$) & \textbf{Correct}       \\
   \vspace{-0.1cm}
     &  &   &       \\
     \hline
   \vspace{-0.1cm}
        &  &   &       \\
 $X_6|X_1,X_2,X_5$&  Laplace$\Bigl[e^{0.03x_1+0.02x_2+0.01x_2^2+0.02x_5},1\Bigl]$&    Laplace$\Big[e^{0.03x_1+0.02x_2+0.02x_5},1\Bigl]$&     No  \\
    \vspace{-0.1cm}
 &&&\\
 $X_1,X_2,X_5$& $N\left[\left(\begin{array}{c}
10\\
15\\
11
\end{array}\right),\left(\begin{array}{ccc}
4 & 0.5 & 0\\
0.5 & 3 & 1\\
0 & 1 & 5
\end{array}\right)\right]$ & $N\left[\left(\begin{array}{c}
10\\
15\\
11
\end{array}\right),\left(\begin{array}{ccc}
4 & 0.5 & 0\\
0.5 & 3 & 1\\
0 & 1 & 5
\end{array}\right)\right]$  &    Yes     \\
   \vspace{-0.1cm}
 &&&\\
 $X_4|X_3$& Exponential$\Bigl(\frac{1}{x_3}\Bigl)$ &   Exponential$\Bigl(\frac{1}{x_3}\Bigl)$  &     Yes    \\
    \vspace{-0.1cm}
  &&&\\
  $X_3$& Exponential$(1)$ &  Exponential$(0.9)$ &  No     \\
     \vspace{-0.1cm}
   &&&\\
$X_7$& $T_3$ & Cauchy$(0,1)$  &  No      \\
 &&&\\
\noalign{\global\arrayrulewidth0.05cm}
 \hline
 \noalign{\global\arrayrulewidth0.05pt}
\end{tabular}
\caption{True and postulated model for Example II. The last column highlights where mismodeling occurs. }
\label{mismodeling}}
\end{table}
\end{center}
Because of condition \eqref{C1}, Theorem \ref{anova_theo}  holds only for random sub-vectors of $\bm{X}$ whose Rosenblatt transform $u_q$ includes all the conditioning, from the higher to the lower, necessary to recover $g_{q}\bigl(\bm{x}_q\bigl)$.  To some extent, this condition can be seen as the iGOF counterpart of  the marginality principle advocated by \citet{nelder} in the context of ANOVA, and which consists in  taking into account of the hierarchy  of the main effects and interactions  in a given model.

Similarly to the ANOVA, Theorem \ref{anova_theo} allows us to construct an iGOF-diagnostic table to identify the source of mismodeling for a given random vector $\bm{X}$ and its components. Below we show how this can be done in practice for the case of a $7$-dimensional random vector.

\emph{\textbf{Example II.}} We consider a sample of $n=5000$ observations from a random vector $\bm{X}=(X_1,\dots,X_7)$ with components distributed as summarized in the second column of Table \ref{mismodeling}.    Table \ref{anova} collects the results obtained by applying Theorem \ref{anova_theo} to test the validity of the models specified for different sub-vectors of $\bm{X}$. The overall deviance test   is reported in the first row and correctly reject the null model.
Similarly, the test in the second row, rejects the hypotheses that the vector $(X_1,X_2,X_5,X_6)$ is modelled correctly, and fails to rejects the model for $(X_1,X_2,X_5)$. 

  \begin{minipage}{\textwidth}
  \begin{minipage}[b]{0.49\textwidth}
    \centering
\fontsize{9}{9}\selectfont{
 \centering
\begin{tabular}{|c|c|c|}
\noalign{\global\arrayrulewidth0.05cm}
 \hline
 \noalign{\global\arrayrulewidth0.05pt}
$\bm{X}_q$  & \textbf{df}& \textbf{(Adjusted)}  \\
  &&\textbf{ p-value }    \\
     \hline
 $\bm{X}$& 16383  &  $<10^{-130}$        \\
 $(X_1,X_2,X_5,X_6)$& 256   & $<10^{-130}$        \\
 $(X_1,X_2,X_5)$&63   &$1$     \\
 $(X_3,X_4)$&  15 & $8.467\cdot 10^{-08}$      \\
  $X_3$&3  &$1.525\cdot 10^{-13}$        \\
$X_7$& 3   & $2.599\cdot 10^{-122}$    \\
\noalign{\global\arrayrulewidth0.05cm}
 \hline
 \noalign{\global\arrayrulewidth0.05pt}
\end{tabular}
\captionof{table}{iGOF-diagnostic table. The third column reports the (post-selection adjusted) deviance p-values  in \eqref{test3} with $\bm{X}_q$ specified as in the first column. The second column corresponds to the degrees of freedom used in the calculation of the p-value, namely, $M_{q}$.  }
\label{anova}}
  \end{minipage}
  \hfill
  \begin{minipage}[b]{0.49\textwidth}
    \centering
   \centering
\includegraphics[width=55mm ]{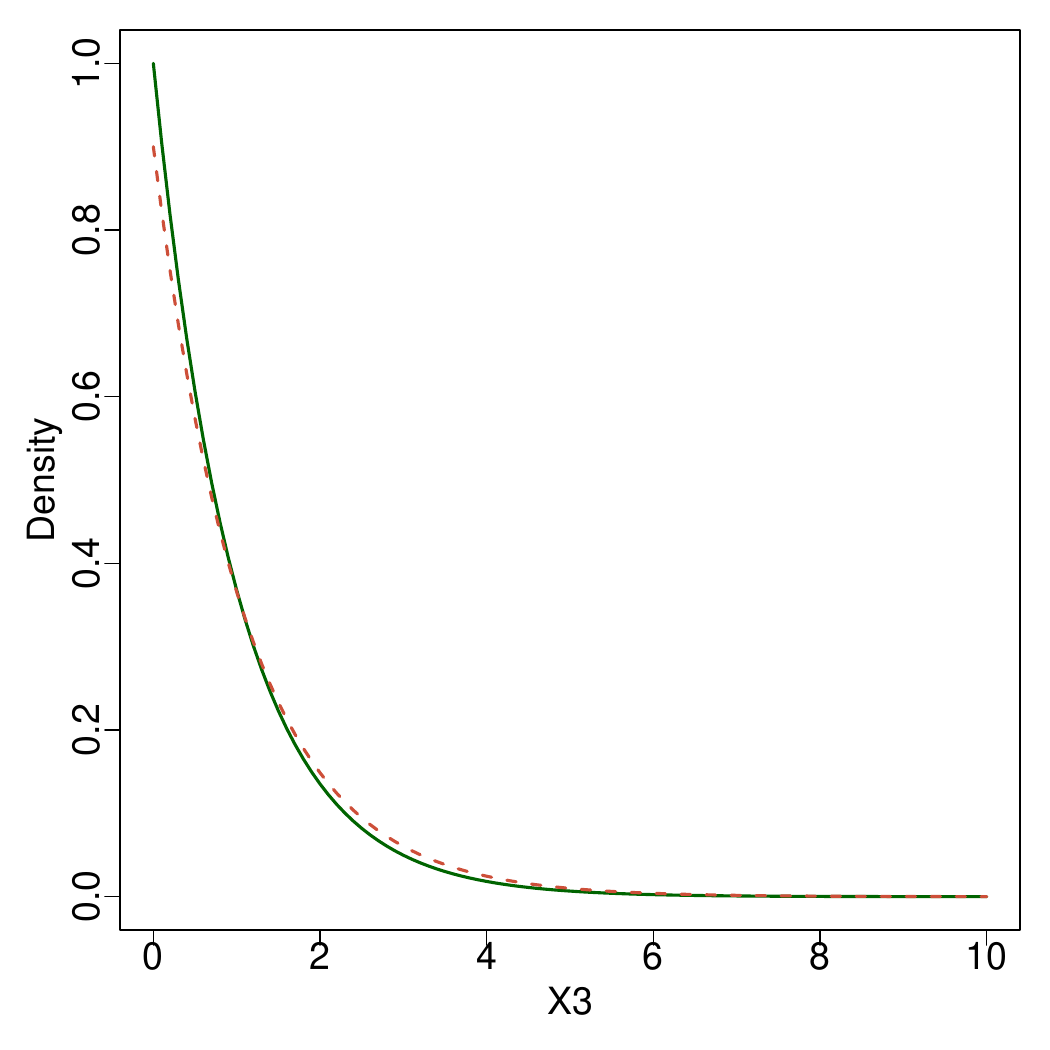} \\
\captionof{figure}{Comparing the postulated (red dashed line) and the true model (green solid line) of $X_3$. }
\label{figexp}
    \end{minipage}
  \end{minipage}\\
\vspace{0.5cm}
  \begin{table}[!h]
   \fontsize{8}{8}\selectfont{
 \centering
 \begin{tabular}{|c|cccccc|}
\noalign{\global\arrayrulewidth0.05cm}
 \hline
 \noalign{\global\arrayrulewidth0.05pt}
       &  &   & & &   &        \\
\multirow{2}{*}{$\bm{X}_q$}  &\multicolumn{6}{c|}{\textbf{Sample size ($n$)}}    \\
      &  &   & & &   &        \\
 &$500$ & $1000$   &$2000$&$3000$& $5000$ &    $10,000$     \\
     \hline
      &  &   & & &   &        \\
$\bm{X}$& 1 &   1  &   1    & 1 & 1   &  1   \\
$(X_1,X_2,X_5,X_6)$& 1 & 1        & 1& 1    &  1  & 1   \\
$(X_1,X_2,X_5)$& 0 & 0 &     0 &  0  &   0 & 0   \\
 $(X_3,X_4)$& 0.0069   & 0.0342 & 0.2384  &  0.5939&  0.9615   &1      \\
 &   ($\pm$0.0008) &   ($\pm$0.0018)&   ($\pm$0.0049)  &  ($\pm$0.0049) &   ($\pm$ 0.0019)  &     \\
 $X_3$  & 0.2360  &0.5560     & 0.9153   &0.9868 &0.9999  &1    \\
  &  ($\pm$0.0043)&   ($\pm$0.0050)   &   ($\pm$0.0028) &  ($\pm$0.0011)&  ($\pm$ 0.0001) &    \\
 $X_7$& 1  &1      &1& 1& 1       &1   \\
       &  &   & & &   &        \\
\noalign{\global\arrayrulewidth0.05cm}
 \hline
 \noalign{\global\arrayrulewidth0.05pt}
\end{tabular}
\caption{Performance of the iGOF-diagnostic analysis for different sample sizes.   For values different from zero and one the Monte Carlo errors $(\pm SE)$ are also reported. The significance level considered is $\alpha=0.05$.}
\label{simul}}
\end{table}
\begin{figure*}[htb]
\begin{tabular*}{\textwidth}{@{\extracolsep{\fill}}@{}c@{}c@{}}
\includegraphics[width=72mm ]{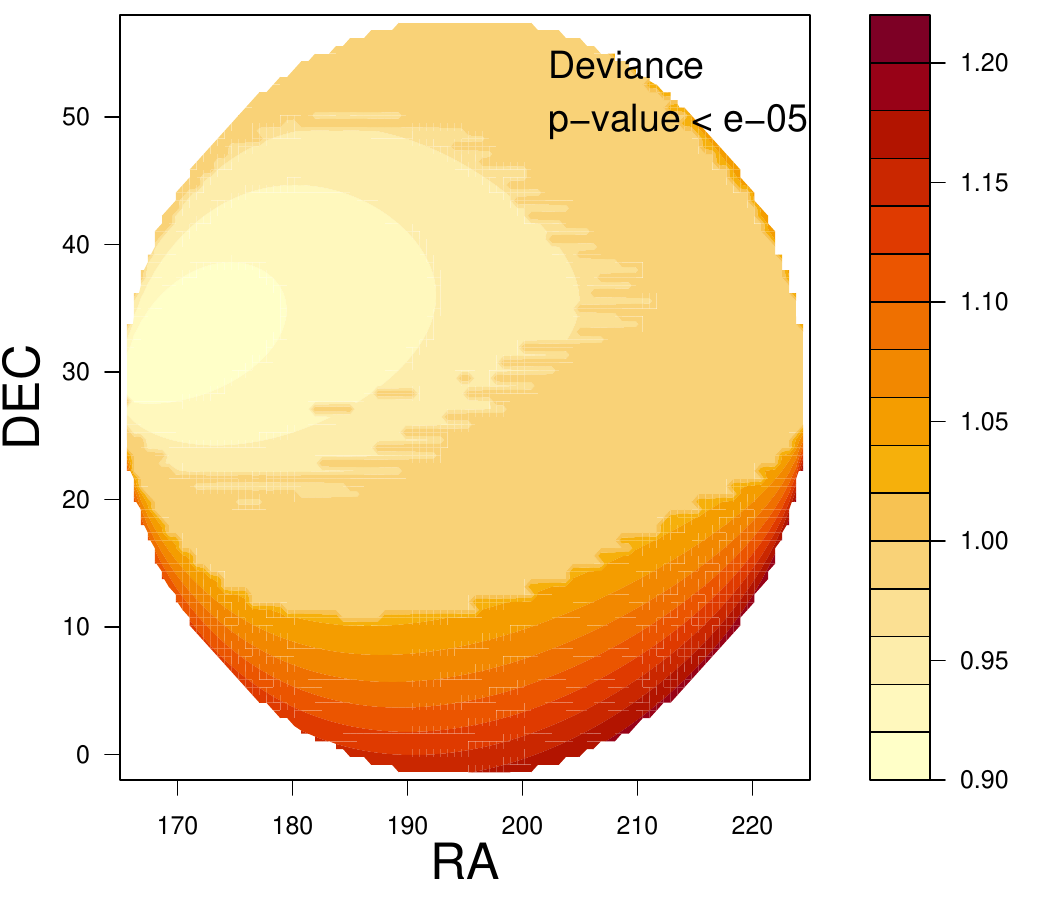}&\includegraphics[width=72mm ]{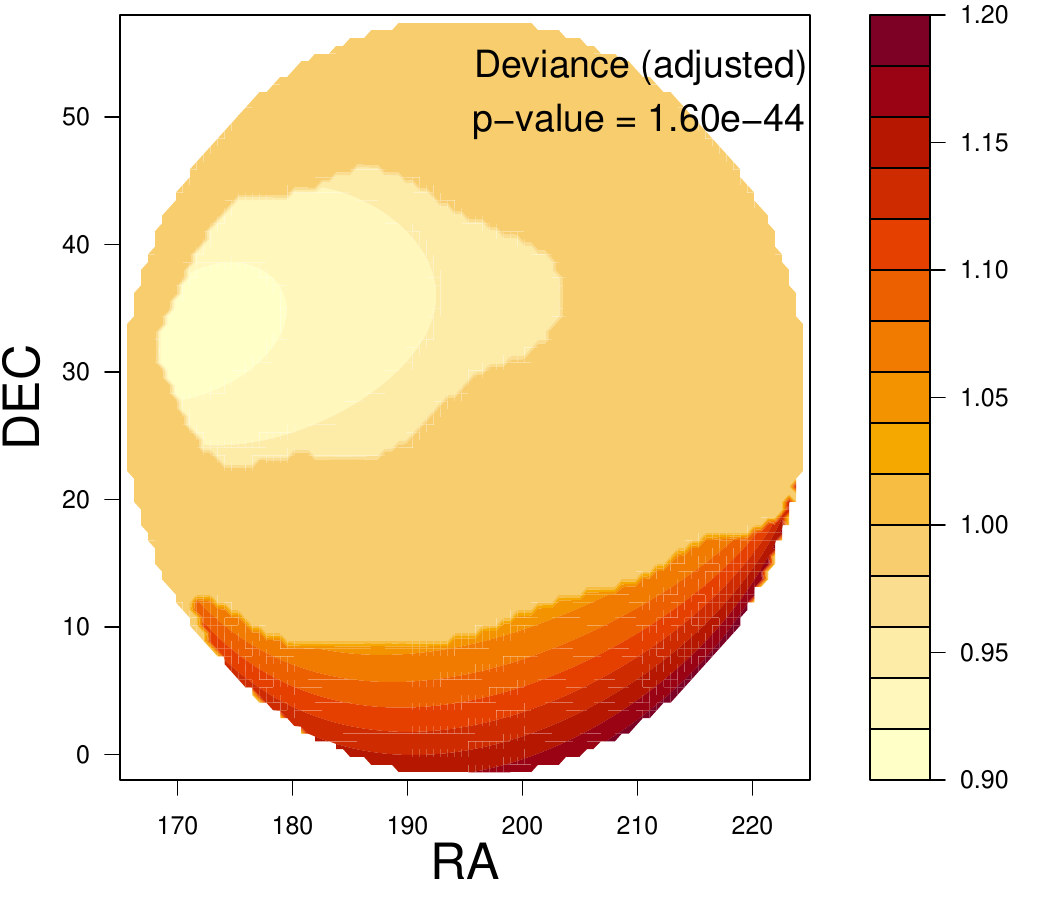} \\
\end{tabular*}
\caption{Simulated and approximated confidence regions for the  Fermi LAT simulation. The left panel corresponds to the (post-selection) confidence regions and deviance p-value   obtained via a simulation of size $B=10,000$. The right panel shows to the (post-selection adjusted) confidence regions and deviance  p-value computed as in \eqref{pvalue} and \eqref{CIband}. Darker shades correspond to significant deviations of the estimated likelihood ratio above one. Lighter shades correspond to significant deviations below one. 
}
\label{Fermi}
\end{figure*}

\noindent This aspect is particularly important as it highlights that the mismodeling occurs only with respect to the conditional distribution of $X_6|X_1,X_2,X_5$. The tests in the fourth and fifth row show that the vector $(X_3,X_4)$ has been mismodeled and one source of mismodeling is the marginal of $X_3$. Ultimately, the test for $X_7$ also correctly rejects  the null hypothesis of Cauchy distribution.
Table \ref{simul} collects the results of a simulation obtained by repeating the diagnostic analysis in Table \ref{anova} through a simulation of $B=10,000$ replicates, while considering different sample sizes. Even when the sample size considered is only $500$,  the most prominent deviations are captured with probability one, whereas, the model for $(X_1,X_2,X_5)$ is never rejected. More issues arise in diagnosing mismodeling of $X_3$ and, consequently, $(X_3,X_4)$ for smaller samples. For instance, even when $n=1000$ the power of the procedure in detecting departures of $G_{X_3}$ from $F_{X_3}$ is only $\sim56\%$ and $\sim3\%$ for $(X_3,X_4)$.

It has to be noted, however, that detecting mismodeling of $X_3$ is a particularly challenging task. As shown in Figure \ref{figexp}, the postulated and the true pdf of $X_3$ are very close one-another; this minor differences are further ``diluted'' when considering the joint distribution of $(X_3,X_4)$, since $X_4|X_3$ is correctly specified. Nevertheless,  such minor deviations are detected with high power for larger sample sizes.

\section{A diagnosis of background mismodeling}
\label{bkg}

When conducting searches for new phenomena, mismodeling of the background distribution  can dramatically compromise 
the sensitivity of the experiment. Specifically, overestimating the background can increase the chances of false negatives. Whereas, underestimating the background may lead to claiming false discoveries. To illustrate how iGOF can be used to understand if and how the postulated background model have been misspecified, we consider a simulated observation by the Fermi Large Area Telescope (LAT) \cite{atwood} obtained with the \emph{gtobssim}  package\footnote{\url{http://fermi.gsfc.nasa.gov/ssc/data/analysis/software}} and previously published in \citet{TOHM}. The simulation includes a 
realistic representations of the instrumental noise of the detector and present backgrounds.

The region of interest corresponds to a disc  in the sky of $30^\circ$ radius and centered at  ($195$ RA, $28$ DEC), where RA and DEC are the coordinates in the sky. Here we assume that, \red{while} the cosmic background is known to follow a uniform distribution over the search area, it is unclear if the instrumental error  is effectively negligible, or if it has a prominent effect on the underlying distribution. Therefore, we set $G_{X_1X_2}$ to be the cdf of a uniform distribution with support $\mathcal{X}_1\times\mathcal{X}_2=[165,195]\times[28-\sqrt{30^2-(x-195)^2}, 28+\sqrt{30^2-(x-195)^2}]$ and we proceed by estimating the likelihood ratio via \eqref{jcd_est} over a sample of $n=68658$ observations. Specifically, we set $m_1=m_2=4$ and we select the components of $\widehat{\bm{\theta}}$ via the BIC criterion in \eqref{BIC}. The resulting estimate is
\begin{equation}
\label{Fermi_est}
\widehat{d}(x_1,x_2)=1+0.022T_{1}\bigl(G_1(x_1)\bigl)-0.043T_{1}\bigl(G_2(x_2|x_1)\bigl)+0.041T_{2}\bigl(G_2(x_2|x_1)\bigl).
\end{equation} 

In order to assess the significance of the deviations captured by \eqref{Fermi_est}, we compute simultaneous confidence regions and deviance p-values via \eqref{CIband} and \eqref{pvalue}. The results are reported in the right panel of Figure \ref{Fermi}, whereas the left panel shows the confidence regions and deviance p-value obtained via simulation. Similarly to what we have observed for Example I (see Figure \ref{fig2}), despite the approximate confidence bands are more conservative, they still allow to capture the main departures from uniformity.  Indeed, in both cases, we can see that the prominent deviations of the true underlying model from the postulated uniform distribution occur in proximity of low values of $X_2$.   Whereas, at the center-left of the search area, the uniform model significantly underestimates the model inclusive of the instrumental error. Finally, it follows from \eqref{skewG_est} that an updated model background distribution which accounts for these deviations can be constructed as in \eqref{skewG_est} by simply multiplying the uniform pdf by the estimated likelihood ratio in  \eqref{Fermi_est}.

\section{Extensions to the discrete case}
\label{discrete}
The methods discussed so far  focus   on the case where $F$ and $G$ are continuous. However, extensions to the discrete setting can be derived by rewriting   the expansion in \eqref{jcd_rep} through an orthonormal set of functions suitable to model discrete data. This can be done, for instance, by means of the so-called ``LP\footnote{In the \emph{LP} acronym, the letter \emph{L} typically denotes nonparametric methods based on quantiles, whereas \emph{P} stands for polynomials \cite[Supp S1]{LPksamples}.}  score functions'', recently introduced \citep[e.g.,][]{LPksamples} and which can be seen as a generalization of the   Legendre polynomials valid in both  the continuous and discrete setting. 

Specifically, when $p=1$,  a complete orthonormal basis of LP score functions in $L^2(G)$   can be specified by letting the first component to be $T_0\bigl[G(x)\bigl]=1$. Subsequent components $\{T_j\bigl[G(x)\bigl]\}_{j>0}$ are obtained by Gram-Schimidt orthonormalization of powers of
\begin{equation}
\label{T1}
T_1\bigl[G(x)\bigl]=\frac{G_{\text{mid}}(x)-E[G_{\text{mid}}(x)]}{\sqrt{V(G_{\text{mid}}(x))}}=\frac{G(x)-0.5p_{G}(x)-0.5}{\sqrt{[1-\sum_{x\in \mathcal{X}}p_{G}^3(x)]/12}},
\end{equation}
where  $G_{\text{mid}}(x)=G(x)-0.5p_{G}(x)$ is the  \emph{mid-distribution function}, which it has been shown in \citet{parzen2004} to have  mean $0.5$ and variance $[1-\sum_{x\in \mathcal{X}}p_{G}^3(x)]/12$, with $\mathcal{X}$ being the set of distinct points in the support of $X$ and $p_{G}(x)=P(X=x)$ if $X\sim G$. 
Therefore, $T_1\bigl[G(x)\bigl]$ is the standardized mid-distribution and orthonormality of the $T_j\bigl[G(x)]$ functions  in  $L^2(G)$ follows by the first equality in \eqref{T1} and by Gram-Schmidt  process. 

Notice that, for continuous $X$, $G_{\text{mid}}(x)=G(x)$ and $\sum_{x\in \mathcal{X}}p_{G}^3(x)=0$, consequently, the LP score functions reduce to normalized shifted Legendre polynomials. The latter are effectively the result of a Gram-Schmidt orthonormalization applied to powers of $G(x)$. 
Whereas, the LP score functions are obtained  by orthonormalizing powers of the standardized mid-distribution function  with respect to the measure $G$. 

Recall that, in our context, the cdfs $G_d$, $d=1,\dots,p$ are the conditional and marginal distribution functions specified in the Rosenblatt's transform $\bm{G}_R(\bm{x})$. 
Hence, an orthonormal basis in $L^2\bigl(G_d)$ is given by the set of functions $T_{j_d}\bigl[G_d(x_d|\bm{x}_{<d})\bigl]$, $j_d\geq 0$ with $T_{0}\bigl[G_d(x_d|\bm{x}_{<d})\bigl]=1$
and subsequent components

\begin{align}
\label{T0etc}
\quad T_{j_d}\bigl[G_d(x_d|\bm{x}_{<d})\bigl]&=\frac{\text{\r{T}}_{j_d}\bigl[G_d(x_d|\bm{x}_{<d})\bigl]}{\bigl|\bigl|\text{\r{T}}_{j_d}\bigl[G_d(x_d|\bm{x}_{<d})\bigl]\bigl|\bigl|_{G_d}}, \quad \text{for all $j_p\geq 1$, where}\\
\label{Tjd}
\text{\r{T}}_{j_d}\bigl[G_d(x_d|\bm{x}_{<d})\bigl]&=T^{j_d}_{1}\bigl[G_d(x_d|\bm{x}_{<d})\bigl]\\
&-\sum_{k=1}^{{j_p}-1}\bigl<T^{{j_d}}_1\bigl[G_d(x_d|\bm{x}_{<d})\bigl],T_k\bigl[G_d(x_d|\bm{x}_{<d})\bigl]\bigl>_{G_d}T_k\bigl[G_d(x_d|\bm{x}_{<d})\bigl],\\
\bigl<T_1^{j_d}\bigl[G_d(x_d|\bm{x}_{<d})\bigl],&T_{k_d}\bigl[G_d(x_d|\bm{x}_{<d})\bigl]\bigl>_{G_d}=\\&\int T_1^{j_d}\bigl[G_d(x_d|\bm{x}_{<d})\bigl]T_{k_d}\bigl[G_d(x_d|\bm{x}_{<d})\bigl]\text{d}G_d(x_d|\bm{x}_{<d})
\end{align}
 and $||\cdot||_{G_d}=\sqrt{<\cdot,\cdot>_{G_d}}$.
 
 When $p>1$ a suitable tensor basis  in  $L^2(G)$ can then be constructed as in  \eqref{Sjs}. 
$\{T_{j_1,\dots,j_p}(\bm{u})\}_{j_1\dots j_p\geq 0}$. Orthonormality of the $T_{j_1,\dots,j_p}(\bm{u})$ score functions can be verified directly as shown in Section 3 of the Supplementary Material.

\section{Discussion}
\label{conclusions}

This work proposes an informative approach to goodness-of-fit which connects exploratory and confirmatory data analysis to study  multivariate distributions. By transforming the likelihood ratio on the unit cube, confidence regions can be constructed as in Corollary \ref{naive2} to identify regions of the supportwhere significant deviations occurs. \red{While} this approach is practical only for problems in at most three dimensions, in more dimensions a detailed diagnosis of mismodeling can be achieved by means of the   iGOF-diagnostic analysis proposed in Section \ref{diagnostics}. These tools can be used to directly address Q1 in Section \ref{intro}.
For instance, given the panacea of theories available on the nature of dark matter, experimentalists aiming to detect it often face the dilemma of selecting which of the tens of theoretical models (mainly non-nested) available should be tested \citep[e.g.,][]{pat}. If one was to test it using the procedure discussed in this paper, even when a given model is rejected, it is possible to gain further insight on the shape of the departure of the true data distribution and the null model and ultimately use such information to ``rule out'' other models which would be inconsistent with such deviation. 

Moreover, as we aimed for when formulating Q2 in Section \ref{intro}, the true probability function of the data can be estimated semi-parametrically via  \eqref{skewG_est}, while assessing the validity of the model postulated by the scientists. Interestingly,  the resulting estimate incorporates the knowledge carried by the hypothesized model and thus,  it provides a data-driven update for it in the direction of the true distribution of the data. 

Despite the usefulness of the methods presented here in applied settings, and in the physical sciences in particular (e.g., Section \ref{bkg}), they are not exempt from limitations. For instance, several problems in physics and astronomy, often involve no more than 8 or 10 dimensions and/or can be reduced to 2D planes \citep[e.g.,][]{aprile}. In this context, choosing  $m_d$ equal to $3$ or $4$ for all $d=1,\dots,p$, is often sufficient to avoid overfitting and, eventually, lack of power by implementing adequate model selection strategies  and for sufficiently large samples (see Sections \ref{inference} and \ref{diagnostics}). In more dimensions, however, the method suffers from the curse of dimensionality \citep[e.g.][]{friedman}, as the size of the LP tensor basis   increases exponentially fast with $p$.  In this context, a regularized solution could be particularly valuable \citep[see for instance][]{signoretto} when analyzing, for instance,   data coming from large astronomical surveys such as the Large Synoptic Survey Telescope (LSST) survey \citep[e.g.,][]{tyson}. Alternatively, if the interest is merely in detecting signals without assuming a specific background model, a data-driven solution for high-dimensional data has been recently proposed by \eqref{mikael}.

Furthermore, the unitary representation of the likelihood ratio  in  \eqref{jcd} relies on the Rosenblatt transform and which can lead to different configurations of $\bm{U}$ and, potentially, different estimators. While this aspect would require adequate treatment on its own, it is worth noting that this problem is essentially the same arising in the context of vine copulas \citep[e.g.,][]{Nagler} and for which adequate model selection procedures exists  \citep[e.g.,][]{panagiotelis,dissmann}.

Finally, the inferential procedures presented here extend classical smooth tests to the multivariate setting and allow us to visualize graphically the departure of $F$ from $G$ and study their substructures. Despite this article focuses on simple null hypothesis, that is, the postulated model is assumed to be fully specified,  
 classical results on smooth tests \citep[e.g.,][Sec 4.2.2.3 and 5.2.2.3 ]{thas} can be used to show to derive asymptotic tests in the parametric setting. Unfortunately, however, the  asymptotic approximations are known to be rather slow in the parametric case. Therefore, in practical applications, when $G$ depends on unknown parameters it is recommended to perform inference by means of the parametric bootstrap and which has been shown by \citet{babu} to be consistent also in the multivariate setting.

\appendix{\Large{\textbf{Appendix}}}

\begin{table}[!h]
\fontsize{9}{9}\selectfont{
 \centering
\begin{tabular}{|r|l|}
\noalign{\global\arrayrulewidth0.05cm}
 \hline
 \noalign{\global\arrayrulewidth0.05pt}
  & \\
Symbol  & Description   \\
 \hline
   & \\
$\bm{X}=(X_1,\dots,X_p)$   & Random vector of components\\ 
& $X_d,d=1,\dots,p$\\
$\mathcal{X}$   & Support of $\bm{X}$\\
$F,f$   & True cdf and density of $\bm{X}$ \\
$G,g$   & Postulated cdf and density of $\bm{X}$ \\
$G_d,G^{-1}_d,g_d$ & Conditional cdf, quantile  \\
&function and density of $X_d$\\
$\bm{U}=\bm{G}_R(\bm{X})$ & Rosenblatt's transform of $\bm{X}$\\
$\bm{G}^{-1}_R(\bm{U})=\bm{X}$   &   Inverse of the Rosenblatt's \\
&transform\\
$\bm{x}=(x_1,\dots,x_p)=\bm{G}^{-1}_R(\bm{u})$ & Realization of $\bm{X}$ \\
&with components $x_d=G^{-1}_d(u_d)$\\
$\bm{u}=(u_1,\dots,u_{p})=\bm{G}_R(\bm{x})$ & Realization of $\bm{U}$,  with \\
&components $u_d=G_d(x_d|\bm{x}_{<d})$\\
$\bm{x}_{<d}=(x_1,\dots,x_{d-1})$ & First $d-1$ components of $\bm{x}$\\
$\bm{u}_{<d}=(u_1,\dots,u_{d-1})$ & First $d-1$ components of $\bm{u}$\\
$d(\bm{u}),\widehat{d}(\bm{u})$ & Likelihood ratio and its estimate\\
$T_{j_d}\bigl[G_d(x_d|\bm{x}_{<d})\bigl]=T_{j_d}(u_d)$ & \red{$j_{d}$-th} normalized shifted Legendre \\
&polynomial in $L^2(G_d)$ and $[0,1]$\\
$T_{k}\bigl[\bm{G}_R(\bm{x})\bigl]=T_k(\bm{u})$ & Tensor product of $T_{j_d}$ functions\\
&with $k\in\mathcal{K}$, in $L^2(G)$ and $[0,1]^p$\\
$\mathcal{K}$ & sets of $p-$tuple $\{j_1\dots j_p\}$, \\
&$\sum_{d}j_d\neq 0$, $j_d=0,\dots,m_d$\\
$|\mathcal{K}|=M$ &Cardinality of $\mathcal{K}$ \\
$\bm{T}(\bm{u})=\bm{T}\bigl[\bm{G}_R(\bm{x})\bigl]$ & $M\times 1$ vector of components \\
&$T_k(\bm{u})=T_k\bigl[\bm{G}_R(\bm{x})\bigl]$\\
$\bm{\theta}$& $M\times 1$ vector collecting the \\
&  coefficients $\theta_k$ \\ 
$\widehat{\bm{\theta}}$& $M\times 1$ vector collecting the \\
&  estimates $\widehat{\theta}_k$\\ 
$D$& Deviance statistics\\
\noalign{\global\arrayrulewidth0.05cm}
 \hline
 \noalign{\global\arrayrulewidth0.05pt}
\end{tabular}
\caption{A summary of the main notation used throughout the paper.}
\label{time}}
\end{table}

\section*{Supplementary Material}{
\textcolor{black}{
The folder \texttt{\emph{Codes\_and\_data}} collects the data used in Section \ref{bkg} and  the codes used  for the analyses  in Figures \ref{fig1}, \ref{fig2}, and \ref{Fermi}. The interested reader is directed to the supplementary file \texttt{\emph{Codes\_description.pdf}} } for a more detailed description of all the  codes and files available.  The file  \texttt{\emph{Supplement\_iGOF.pdf}}  collects the technical proofs and additional results related to Example I.  }

\section*{Acknowledgments}

The author thanks sincerely    G. Jogesh Babu and two anonymous referees  for the
useful suggestions  and comments. Their valuable feedback has led to a substantial improvement of the quality and clarity of the manuscript. \\

{\fontsize{3mm}{3mm}\selectfont{
\setlength{\bibsep}{0pt plus 0.4ex}
\bibliography{biblioLP2}
}}
\end{document}